\documentclass[aps,preprint,amsmath,amssymb,superscriptaddress,nofootinbib]{revtex4}

\usepackage{graphicx}
\begin{document}

\title{Non-standard Higgs couplings in single Higgs boson production at the LHC and future linear collider}

\author{G. Akkaya Sel\c{c}in}
\email[]{akkayag@ankara.edu.tr} \affiliation{Department of Physics,
Faculty of Sciences, Ankara University, 06100 Tandogan, Ankara,
Turkey} \affiliation{Department of Physics, Faculty of Arts and
Sciences, Bitlis Eren University, 13000 Bitlis, Turkey}

\author{\.{I}. \c{S}ahin}
\email[]{inancsahin@ankara.edu.tr}
 \affiliation{Department of
Physics, Faculty of Sciences, Ankara University, 06100 Tandogan,
Ankara, Turkey}

\begin{abstract}
We investigate the potential of single Higgs boson photoproduction
at the LHC and at $e\gamma$ mode of future linear $e^-e^+$ collider
to probe non-standard $HZ\gamma$ and $H\gamma\gamma$ couplings. We
consider the semi-elastic production process $pp\to p\gamma p\to
pHqX$ at the LHC where $q$ represents the quarks and $X$ represents
the remnants of one of the initial protons. We also study the single
Higgs production through $\gamma e\to He$ in the $e\gamma$ collision
at the future linear collider. We perform a model-independent
analysis and obtain the sensitivity bounds on the non-standard Higgs
couplings for both colliders. We compare the capability of single
Higgs photoproduction process at these two colliders to probe
non-standard Higgs couplings.

\end{abstract}


\maketitle

\section{Introduction}

The Higgs boson predicted by Standard Model (SM) of particle physics
was discovered by ATLAS and CMS Collaborations at Large Hadron
Collider (LHC) \cite{Aad:2012tfa,Chatrchyan:2012xdj}. After its
discovery, intense experimental studies have been carried out to
reveal its properties and couplings to other SM particles
\cite{Aad:2013wqa,Chatrchyan:2012jja,Khachatryan:2014kca}. Precise
determination of the Higgs couplings will either confirm the gauge
structure of SM, or provide signal of new physics beyond SM. In this
paper we investigate the non-standard couplings of the Higgs to
gauge bosons $Z$ and $\gamma$ through semi-elastic production of the
Higgs boson at the LHC and single production at $e\gamma$ mode of
future linear $e^-e^+$ collider. These production processes are
electroweak in nature and provide clean channels with respect to
deep inelastic proton-proton collision at the LHC. Therefore, they
can be used to perform precision measurements of the Higgs
couplings.

Non-standard Higgs couplings to gauge bosons have been constrained
through several Higgs decay processes at the LHC
\cite{Corbett:2012dm,Corbett:2012ja,Masso:2012eq,Aad:2013wqa,Banerjee:2013apa,Khachatryan:2014kca}.
There are also experimental constraints obtained from electroweak
precision measurements at LEP and Tevatron
\cite{Corbett:2012dm,Corbett:2012ja,Masso:2012eq,GutierrezRodriguez:2010ri,Group:2012zca}.
One way to examine non-standard Higgs couplings is to employ the
effective lagrangian method. In this method any contribution coming
from new physics beyond SM is described by higher dimensional
operators. These higher dimensional operators are added to the SM
lagrangian and inversely proportional to some powers of $\Lambda$
which is called the scale of new physics. In this paper we analyze
non-standard $HZ\gamma$ and $H\gamma\gamma$ couplings in a model
independent way by means of the effective lagrangian formalism of
Refs.\cite{Buchmuller:1985jz,Leung:1984ni,DeRujula:1991ufe,Hagiwara:1993ck,GonzalezGarcia:1999fq,Corbett:2012dm,Corbett:2012ja,Masso:2012eq}.
There are five $C$ and $P$ even dimension-6 operators which modify
the Higgs boson couplings to $Z$ and $\gamma$ bosons
\cite{Buchmuller:1985jz,Leung:1984ni,DeRujula:1991ufe,Hagiwara:1993ck,GonzalezGarcia:1999fq,Corbett:2012dm,Corbett:2012ja,Masso:2012eq}:
\begin{eqnarray}
\label{effectiveoperators} {\cal O}_{WW}=&&\Phi^\dag\
{\hat{W}}_{\mu\nu}
{\hat{W}}^{\mu\nu}\Phi \nonumber\\
{\cal O}_{BB}=&&\Phi^\dag\ {\hat{B}}_{\mu\nu} {\hat{B}}^{\mu\nu}\Phi\nonumber\\
{\cal O}_{BW}=&&\Phi^\dag\ {\hat{B}}_{\mu\nu} {\hat{W}}^{\mu\nu}\Phi\\
{\cal O}_{W}=&&(D_{\mu}\Phi)^\dag {\hat{W}}^{\mu\nu} (D_{\nu}\Phi)\nonumber\\
{\cal O}_{B}=&&(D_{\mu}\Phi)^\dag {\hat{B}}^{\mu\nu}
(D_{\nu}\Phi)\nonumber
\end{eqnarray}
where $\Phi$ is the scalar doublet, $D_\mu$ is the covariant
derivative, $\hat W_{\mu\nu}=i\frac{g}{2}(\vec \sigma\cdot \vec
W_{\mu\nu})$ and $\hat B_{\mu\nu}=i\frac{g'}{2}B_{\mu\nu}$. Here $g$
and $g'$ are the $SU(2)_L$ and $U(1)_Y$ gauge couplings. The field
strength tensors $W^i_{\mu\nu}$ and $B_{\mu\nu}$ belong to $SU(2)_L$
and $U(1)_Y$ gauge groups respectively. The SM lagrangian is then
modified by the following dimension-6 effective lagrangian:
\begin{eqnarray}
\label{Leff1} {\cal L}_{eff}=\sum_n \frac{f_n}{\Lambda^2}{\cal
O}_{n}
\end{eqnarray}
where $f_n$ denote the non-standard couplings and $\Lambda$ is the
scale of new physics. After symmetry breaking, the effective
lagrangian in Eq.(\ref{Leff1}) give rise to the following $HZ\gamma$
and $H\gamma\gamma$ interactions \cite{Corbett:2012dm}:
\begin{eqnarray}
\label{Leff2}{\cal{L}}_{eff}=g_{H\gamma\gamma}HA_{\mu\nu}A^{\mu\nu}
+{g_{HZ\gamma}^{(1)}}A_{\mu\nu}Z^{\mu}\partial^\upsilon
H+{g_{HZ\gamma}^{(2)}}HA_{\mu\nu}Z^{\mu\nu}.
\end{eqnarray}
where $V_{\mu\nu}=\partial_\mu V_\nu-\partial_\nu V_\mu$ with $V=A$
and $Z$ field. The non-standard couplings $g_{H\gamma\gamma}$,
${g_{HZ\gamma}^{(1)}}$ and ${g_{HZ\gamma}^{(2)}}$ are related to the
couplings $f_n$ appearing in the effective lagrangian (\ref{Leff1})
before symmetry breaking as
\begin{eqnarray}
\label{gHgg}g_{H\gamma\gamma}=-\left(\frac{gM_{W}}{\Lambda^2}\right)
\frac{s^2(f_{BB}+f_{WW}-f_{BW})}{2}\\
\label{gHZg}{g_{HZ\gamma}^{(1)}}=\left(\frac{gM_{W}}{\Lambda^2}\right)
\frac{s(f_{W}-f_{B})}{2c}\\
\label{gHZg}{g_{HZ\gamma}^{(2)}}=\left(\frac{gM_{W}}{\Lambda^2}\right)
\frac{s[2s^2f_{BB}-2c^2f_{WW}+(c^2-s^2)f_{BW}]}{2c}
\end{eqnarray}
where $s=\text{sin}\theta_W$, $c=\text{cos}\theta_W$, $\theta_W$ is
the Weinberg angle and $M_W$ is the mass of the $W$ boson. In the
calculations presented in this paper the energy scale of new physics
is taken to be $\Lambda=1\text{TeV}$. The effective operators in
(\ref{effectiveoperators}) contribute also $HZZ$ and $HWW$
couplings. Since the processes that we consider in this paper do not
contain these couplings, we do not present the contributions coming
from effective lagrangian (\ref{Leff1}) to $HZZ$ and $HWW$. The
effective operator ${\cal O}_{BW}$ modifies also the $W^3-B$ mixing.
It is stringently restricted by the electroweak precision
measurements \cite{DeRujula:1991ufe,Hagiwara:1993ck,Alam:1997nk}.
Therefore, during the analysis we set $f_{BW}=0$ and consider the
contributions from four couplings $f_{WW}$, $f_{BB}$, $f_{W}$ and
$f_{B}$. For the purpose of simplicity, we will consider the
following six different new physics scenarios:
\begin{eqnarray}
&&\underline{Scenario\; I}: f_{B}=f_{W}=0\;,\; f_{BB}=f_{WW} \nonumber \\
&&\underline{Scenario\; II}: f_{B}=-f_{W}\;,\; f_{BB}=f_{WW}=0 \nonumber \\
&&\underline{Scenario\; III}: f_{B}=f_{W}=0\;,\; f_{BB}=-f_{WW} \nonumber \\
&&\underline{Scenario\; IV}: f_{B}=f_{W}=0\;,\;
f_{WW}=\tan^2{\theta_W}f_{BB} \nonumber \\
&&\underline{Scenario\; V}: f_{W}=f_{WW}=0 \nonumber \\
&&\underline{Scenario\; VI}: f_{B}=f_{BB}=0 \nonumber
\end{eqnarray}
In scenarios I$-$IV we impose three constraints and therefore we
have one free parameter. On the other hand, in scenarios V and VI
two constraints are imposed and two parameters remain free. Here we
should note the following important point: In this paper, we employ
the set of bosonic operators in the
Hagiwara-Ishihara-Szalapski-Zeppenfeld (HISZ) basis
\cite{Hagiwara:1993ck}. The operators ${\cal O}_{W}$ and ${\cal
O}_{B}$ do not appear in the Warsaw basis \cite{Grzadkowski:2010es}.
They could be translated into other operators, including ${\cal
O}_{WW},{\cal O}_{BW},{\cal O}_{BB}$ and other dimension-6
operators. Therefore all five operators given in
Eq.(\ref{effectiveoperators}) are not independent. In scenarios I,
III and IV we ignore the contributions from ${\cal O}_{W}$ and
${\cal O}_{B}$ operators which are absent in the Warsaw basis. In
scenarios II, V and VI we consider the contributions from these
operators. However, we consider at most two of the couplings as
independent parameters. Therefore, our scenarios do not overwhelm
the degrees of freedom in the effective lagrangian.


$HZ\gamma$ and $H\gamma\gamma$ interactions do not appear in the SM
at the tree-level. However, they receive contributions at one-loop
level. One-loop contributions to these interactions can be
approximated to the following effective lagrangian
\cite{DynamicsofSM,Farina:2015dua}:
\begin{eqnarray}
\label{loop}
{\cal{L}}_{eff}=g^{(SM)}_{H\gamma\gamma}HA_{\mu\nu}A^{\mu\nu}
+{g^{(SM)}_{HZ\gamma}}HA_{\mu\nu}Z^{\mu\nu}
\end{eqnarray}
where, $g^{(SM)}_{H\gamma\gamma}=\frac{2\alpha}{9\pi\nu}$ and
${g^{(SM)}_{HZ\gamma}}=\frac{\alpha}{4\pi\nu\sin\theta_W}(5.508-0.004i)$.
Here, $\alpha$ is the fine structure constant and $\nu$ is the
electroweak vacuum expectation value.

The semi-elastic single Higgs boson production at the LHC has been
studied in Refs.\cite{Senol:2014naa,Monfared:2016vwr}. However, in
these studies only non-standard $HZ\gamma$ coupling has been taken
into account. In our analysis of semi-elastic Higgs production we
consider both non-standard $HZ\gamma$ and $H\gamma\gamma$ couplings.
We do not assume that $HZ\gamma$ and $H\gamma\gamma$ couplings are
independent from each other. We obtain bounds on $f_n$ couplings of
the operators (\ref{effectiveoperators}) before symmetry breaking
which contribute to both $HZ\gamma$ and $H\gamma\gamma$. The
non-standard Higgs couplings to gauge bosons have also been
investigated at future linear $e^-e^+$ collider and its $e\gamma$
and $\gamma\gamma$ modes
\cite{GonzalezGarcia:1998wn,Banin:1998ap,Hagiwara:2000tk,Ginzburg:2000rm,Han:2000mi,
Barger:2003rs,Han:2005pu,Biswal:2005fh,Choudhury:2006xe,Sahin:2008jc,Sahin:2008qp,Dutta:2008bh,
Biswal:2008tg,Biswal:2009ar,Beneke:2014sba,Cao:2015iua,Kumar:2015eea,Ge:2016zro,Alam:2017hkf}.
The non-standard $HZ\gamma$ and $H\gamma\gamma$ interactions were
investigated through single production process $\gamma e\to He$ in
Refs.\cite{Banin:1998ap,Ginzburg:2000rm}. In
Ref.\cite{Ginzburg:2000rm} the authors analyzed $CP$-odd
interactions which are different from $C$ and $P$ even effective
interactions that we consider. In Ref.\cite{Banin:1998ap} the
authors considered a similar (but not equivalent) effective
lagrangian for the non-standard Higgs interactions. The difference
is that the effective interaction proportional to
${g_{HZ\gamma}^{(1)}}$ (see Eq. (\ref{Leff2})) was omitted in
Ref.\cite{Banin:1998ap}. Another difference between our work and
that of \cite{Banin:1998ap} is that Ref.\cite{Banin:1998ap} was
published long before the discovery of Higgs boson. Therefore, the
authors couldn't perform a detailed statistical analysis considering
the exact value of the Higgs mass. In our analysis of single Higgs
production $\gamma e\to He$, we perform a $\chi^2$ test and estimate
sensitivity of the linear collider based $e\gamma$ collider to
non-standard Higgs couplings for various integrated luminosity
values.

\section{Single Higgs production through photon-proton collision at the LHC}

The LHC is designed as a high-energy proton-proton collider and the
majority of the studies at the LHC focused on deep inelastic
scattering (DIS) processes where both of the colliding protons
dissociate into partons. On the other hand, it was firstly shown
experimentally at the Fermilab Tevatron that complementary to
hadron-hadron collisions, hadron colliders can also be studied as a
photon-photon and photon-hadron collider \cite{cdf1,cdf2,cdf3}.
Recent experimental studies by CMS and ATLAS Collaborations have
verified the existence of such photon-induced reactions at the LHC
\cite{Chatrchyan:2011ci,Chatrchyan:2012tv,Chatrchyan:2013foa,Khachatryan:2016mud,Aaboud:2016dkv}.
It was also shown that these photon-induced processes at the LHC
have a significant potential to probe new physics beyond the SM
\cite{Chatrchyan:2013foa,Khachatryan:2016mud,Aaboud:2016dkv}. The
photon-photon collisions take place when both of the incoming
protons emit quasireal photons. These emitted quasireal photons can
interact mutually and the photon-photon collision occurs as a
subprocess of the proton-proton collision. Similarly when one of the
incoming proton emits a quasireal photon then a photon-proton
collision can occur. These photon-proton collision processes are
sometimes called semi-elastic processes due to their hybrid nature.
Here, the essential point is the distinguishability of such
photon-photon and photon-proton processes from those in which
initial photons are described by propagators. According to
equivalent photon approximation (EPA)
\cite{budnev1975,baur2002,piotrzkowski2001}, emitted photons have a
very low virtuality and up to a high degree of approximation they
are accepted to be real. Furthermore, since the virtuality of the
quasireal photons is very low, photon emitting protons do not
generally dissociate into partons but they remain intact
\cite{rouby,Schul:2011xwa}. After elastic photon emission protons
generally deviate slightly from the direction of beam pipe and
escape from the central detectors without interacting. This causes a
missing energy signature known as the forward large-rapidity gap, in
the corresponding forward region of the central detector
\cite{rouby,Schul:2011xwa,Albrow:2008az}. Moreover, the LHC is
planned to be equipped with very forward detectors which can detect
intact protons escaping from the central detectors
\cite{forward-new1,forward-new2,Tasevsky:2015xya}. The installation
of very forward detectors should allow to separate more easily the
photon-photon and photon-proton processes, where one or both of the
incident protons remain intact
\cite{Albrow:2008pn,Tasevsky:2009zza,Albrow:2010yb,Tasevsky:2014cpa}.
The range of the forward detectors are characterized by the $\xi$
parameter which represents the momentum fraction loss of the proton.
If $\vec p$ represents the initial proton's momentum and
$\vec{p}^{\,\,\prime}$ represents forward proton's momentum after
scattering then, $\xi$ parameter is given by the formula
$\xi\equiv(|\vec{p}|-|\vec{p}^{\,\,\prime}|)/|\vec{p}|$. In this
paper, we will consider a forward detector acceptance range of
$0.015 < \xi < 0.15$
\cite{forward-new1,forward-new2,Tasevsky:2015xya}.

\begin{figure}
\includegraphics[scale=1]{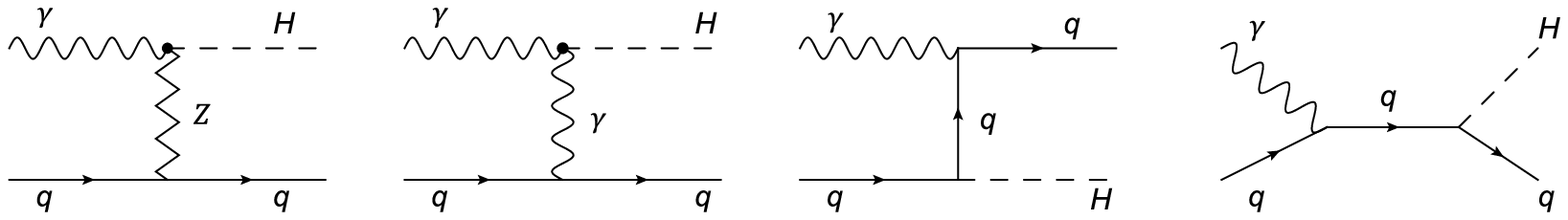}
\caption{Tree-level Feynman diagrams for the subprocess $\gamma q
\to Hq$ \label{fig1}}
\end{figure}

\begin{figure}
\includegraphics[scale=0.6]{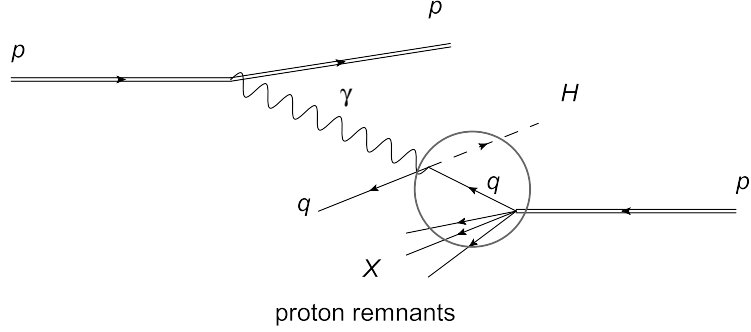}
\caption{The illustration of the process $pp\to p\gamma p\to pHqX$.
\label{fig2}}
\end{figure}

There is an increasing interest in probing new physics through
photon-photon and photon-proton collision at the LHC.
Phenomenological studies on this subject have been growing rapidly
in recent years and cover a wide spectrum of new physics scenarios.
It is impossible to cite all of the references here, but some
representative ones might be Refs.
\cite{Senol:2014naa,Monfared:2016vwr,pheno-1,pheno-2,pheno-3,pheno-4,pheno-5,pheno-6,pheno-7,pheno-8,
pheno-9,pheno-10,pheno-11,pheno-12,pheno-13,pheno-14,pheno-15,pheno-16,pheno-17,pheno-18,pheno-19,pheno-20,
pheno-21,pheno-22,pheno-23,pheno-24,pheno-25,pheno-26,pheno-27} The
semi-elastic single Higgs boson production can be studied through
the process $pp\to p\gamma p\to pHqX$ at the LHC. This process
consists of the subprocesses $\gamma q\to Hq$ where $q$ represents
the quarks. We ignore the top quark distribution and consider 10
independent subprocess for $q=u,d,s,c,b,\bar u,\bar d,\bar s,\bar
c,\bar b$. In the presence of non-standard $HZ\gamma$ and
$H\gamma\gamma$ interactions the subprocess $\gamma q\to Hq$ is
described by the Feynman diagrams given in Fig.\ref{fig1}. The
semi-elastic process $pp\to p\gamma p\to pHqX$ consists of two
different types of proton scattering; elastic photon emission takes
place from one of the initial protons, whereas other initial proton
interact strongly with the emitted photon and undergoes an inelastic
scattering (Fig.\ref{fig2}). Therefore, the cross section for the
semi-elastic process $pp\to p\gamma p\to pHqX$ is obtained by
integrating the cross sections for the subprocesses over the photon
and quark distributions:
\begin{eqnarray}
\label{mainprocess}
 \sigma\left(pp\to p \gamma p\to p H q X\right)=\sum_q\int_{{x_1}_{min}}^{{x_1}_{max}} {dx_1 }\int_{0}^{1}
{dx_2}\left(\frac{dN_\gamma}{dx_1}\right)\left(\frac{dN_q}{dx_2}\right)
\hat{\sigma}_{\gamma q\to H q}(\hat s).
\end{eqnarray}
Here, $\frac{dN_\gamma}{dx_1}$ and $\frac{dN_q}{dx_2}$ are the
equivalent photon and quark distribution functions, respectively.
The quark distribution functions can be evaluated numerically by
using the code MSTW2008 \cite{Martin:2009iq}. In
Eq.(\ref{mainprocess}) the integral variable $x_1$ is the energy
fraction that represents the ratio between the emitted equivalent
photon and initial proton energy. The other variable $x_2$
represents the momentum fraction of the proton's momentum carried by
the quark. The equivalent photon distribution
$\frac{dN_\gamma}{dx_1}$ is given by an analytical expression. We do
not give its explicit form.  Its explicit form can be found in the
literature (for example see \cite{budnev1975} or \cite{pheno-4}). At
the LHC energies where the energy of the incoming proton is much
greater than its mass ($E>>m_p$), the $\xi$ parameter is
approximated as
$\xi\approx\frac{E-E^\prime}{E}=\frac{E_\gamma}{E}=x_1$. Here, $E$
and $E^\prime$ are the energy of the initial and final (scattered)
proton and $E_\gamma$ is the energy of the equivalent photon.
Therefore, the upper and lower limits of the $dx_1$ integration are
determined by the limits of the forward detector acceptance and we
take ${x_1}_{min}=\xi_{min}=0.015$, ${x_1}_{max}=\xi_{max}=0.15$.

\begin{figure}
\includegraphics[scale=1]{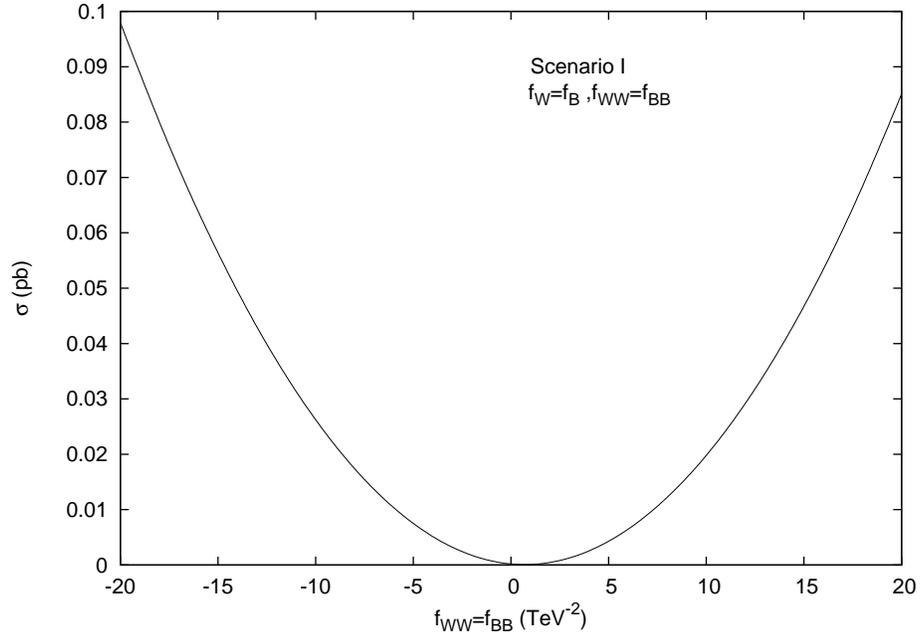}
\caption{The total cross section of the process $pp\to p\gamma p\to
pHqX$ as a function of non-standard Higgs coupling for scenario I.
The center of mass energy of the proton-proton system is taken to be
$\sqrt s=14$TeV. \label{fig3}}
\end{figure}

\begin{figure}
\includegraphics[scale=1]{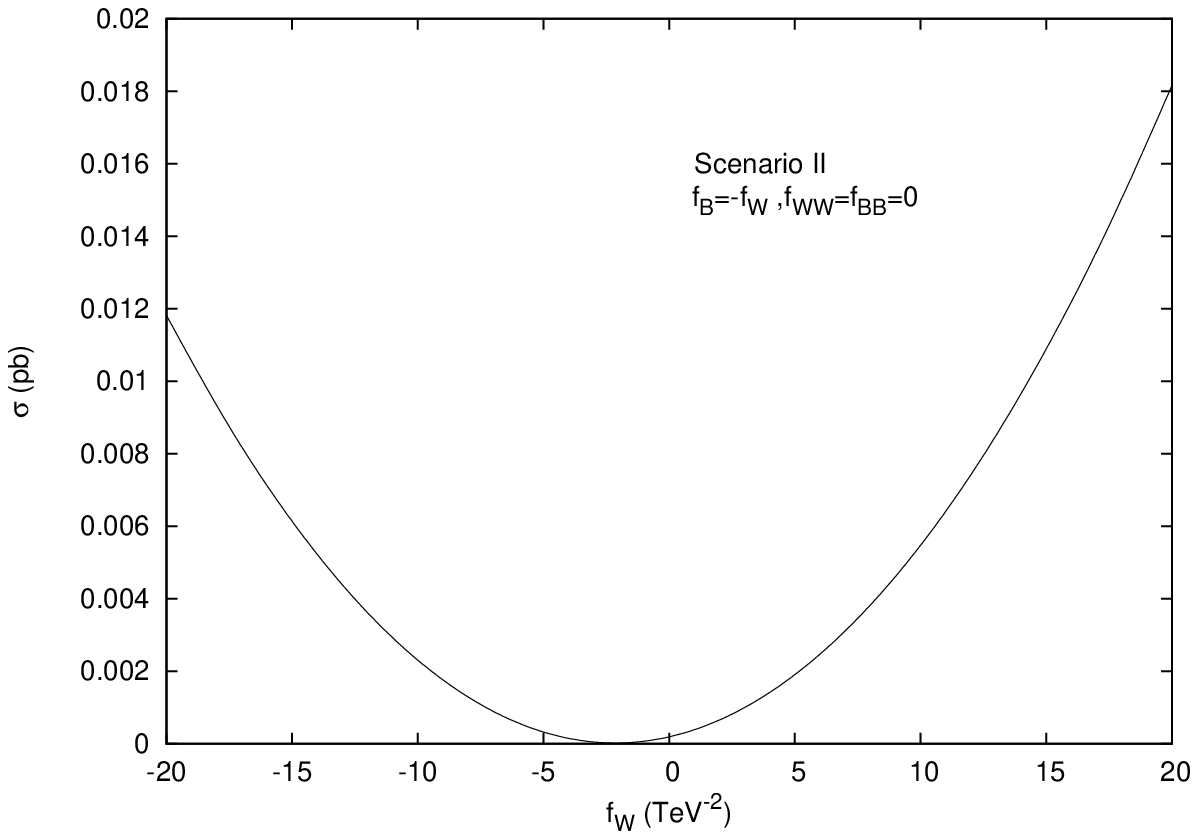}
\caption{The same as Fig.\ref{fig3} but for scenario
II.\label{fig4}}
\end{figure}

\begin{figure}
\includegraphics[scale=1]{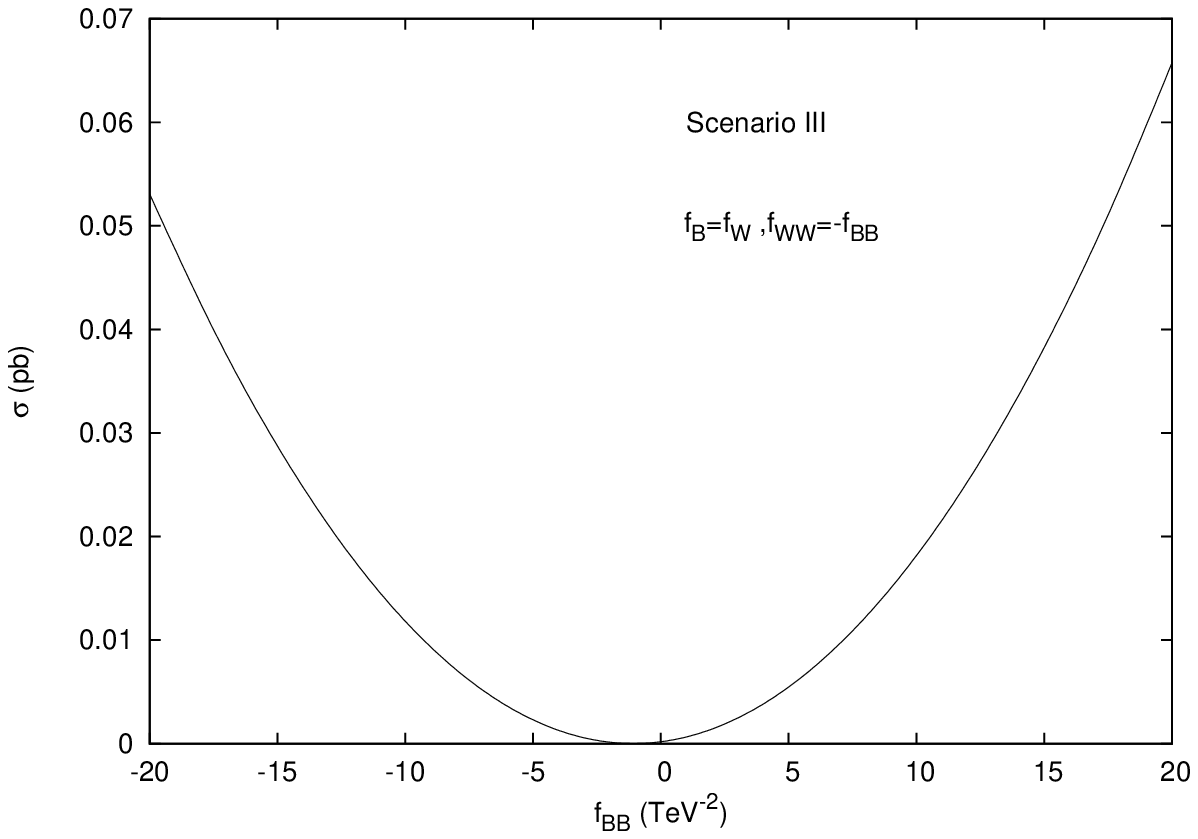}
\caption{The same as Fig.\ref{fig3} but for scenario
III.\label{fig5}}
\end{figure}

\begin{figure}
\includegraphics[scale=1]{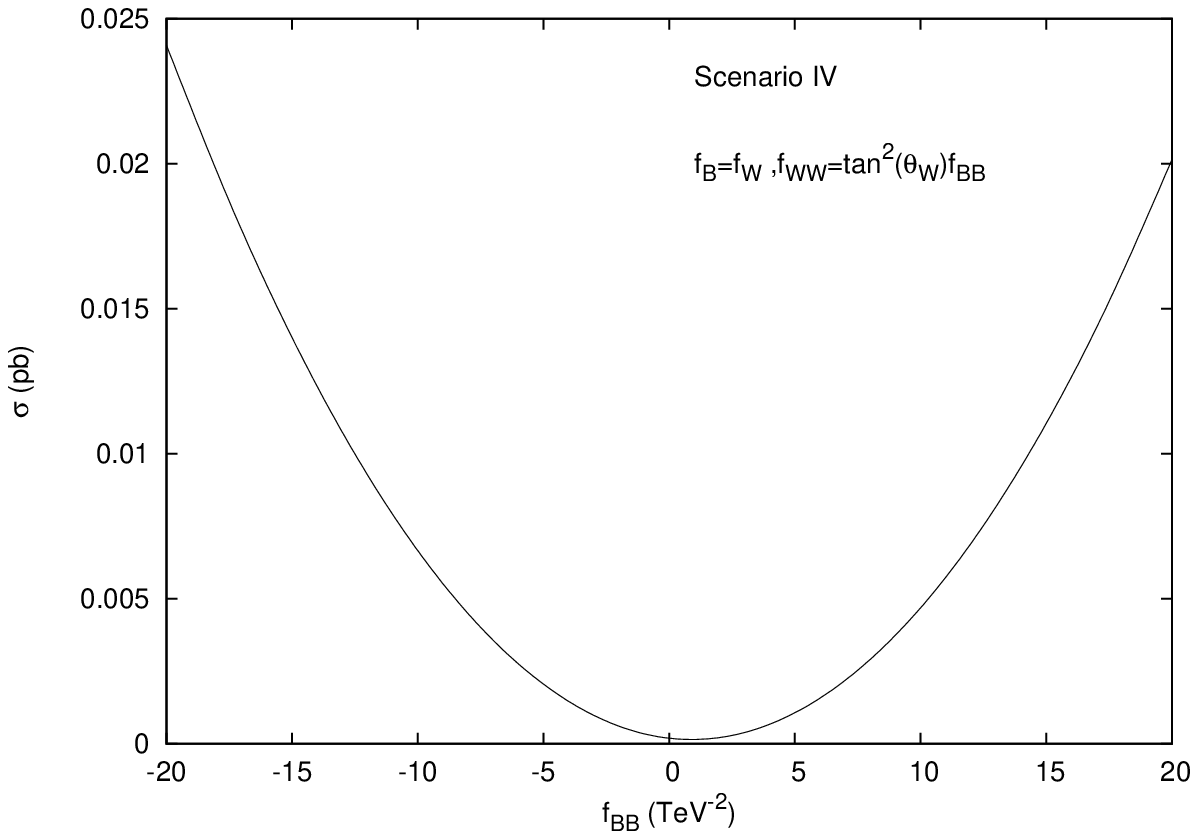}
\caption{The same as Fig.\ref{fig3} but for scenario
IV.\label{fig6}}
\end{figure}

In Figs.\ref{fig3}-\ref{fig6}, we plot the total cross section of
the process $pp\to p\gamma p\to pHqX$ as a function of non-standard
Higgs couplings for scenarios I-IV. In addition to new physics
contributions we have also considered the effective lagrangian
(\ref{loop}) that contains SM one-loop contributions. For a concrete
result we have obtained $95\%$ confidence level (C.L.) bounds on
non-standard couplings using the simple $\chi^2$ criterion. The
$\chi^2$ function is given by
\begin{eqnarray}
\chi^{2}=\left(\frac{N_{NS}-N_{SM}}{N_{SM} \,\, \delta}\right)^{2}
\end{eqnarray}
where, $N_{NS}$ is the number of events containing both new physics
and SM contributions, $N_{SM}$ is the number of events expected in
the SM and $\delta=\frac{1}{\sqrt{N_{SM}}}$ is the statistical
error. The number of events has been calculated considering the
$H\to b \bar b$ decay of the Higgs boson as the signal. Hence, we
assume that $N_{NS(SM)}=E\times S\times  L_{int} \times
\sigma_{NS(SM)}\times BR$ where, $E$ is the b-tagging efficiency,
$S$ is the survival probability factor, $L_{int}$ is the integrated
luminosity and $BR$ is the branching ratio for $H\to b \bar b$.
$\sigma_{SM}$ represents the SM cross section and $\sigma_{NS}$
represents the cross section containing both new physics and SM
contributions. We have taken into account a b-tagging efficiency of
$E=0.6$, survival probability factor of $S=0.7$ and branching ratio
of $BR=0.6$. The survival probability factor of $0.7$ was proposed
for the single W boson photoproduction
\cite{deFavereaudeJeneret:2009db,Khoze:2002dc}. We assume that same
survival probability factor is valid for our process. Although the
b-tagging efficiency is not constant but depends on many different
parameters such as the jet transverse momentum, the algorithm used
in the detector, etc. we assume a constant b-tagging efficiency of
$0.6$. According to experimental works a constant average value of
0.6 for b-tagging efficiency is reasonable \cite{Aad:2015ydr}. We
have also placed a pseudorapidity cut of $|\eta|<2.5$ for final
state particles. There are background processes which contribute to
the same final state. The background processes consist of the SM
subprocesses that contribute to $pp\to p\gamma p\to p b \bar b q X$.
There are totally 18 background subprocess of the type $\gamma q\to
k, b,\bar b$ where, $q=u,d,s,c,b,\bar u,\bar d,\bar s,\bar c,\bar b$
and $k=u,d,s,c,b,t,\bar u,\bar d,\bar s,\bar c,\bar b,\bar t$
quarks. The background contributions have been calculated by using
CalcHEP 3.6.20 \cite{Belyaev:2012qa}. The determination of an
on-shell Higgs boson with mass approximately 125 GeV requires an
invariant mass measurement of the final-state $b \bar b$ pairs. If
we impose a cut and demand that the invariant mass of the $b \bar b$
pairs is in the interval $120\;\text{GeV}<M_{b\bar
b}<130\;\text{GeV}$ then the background cross section is reduced
considerably and gives $\sigma_{\text{background}}=0.05\;\text{pb}$.
Since the background contribution cannot be discerned from Higgs
production cross section, during statistical analysis we add the
background contribution to the SM cross section and assume that
$\sigma_{NS(SM)}=\sigma(pp\to p\gamma p\to p H q
X)_{NS(SM)}+\frac{1}{BR}\times\sigma_{\text{background}}$. Here, the
factor $\frac{1}{BR}$ is used to cancel out the branching ratio in
$N_{NS(SM)}$.

\begin{figure}
\includegraphics[scale=1]{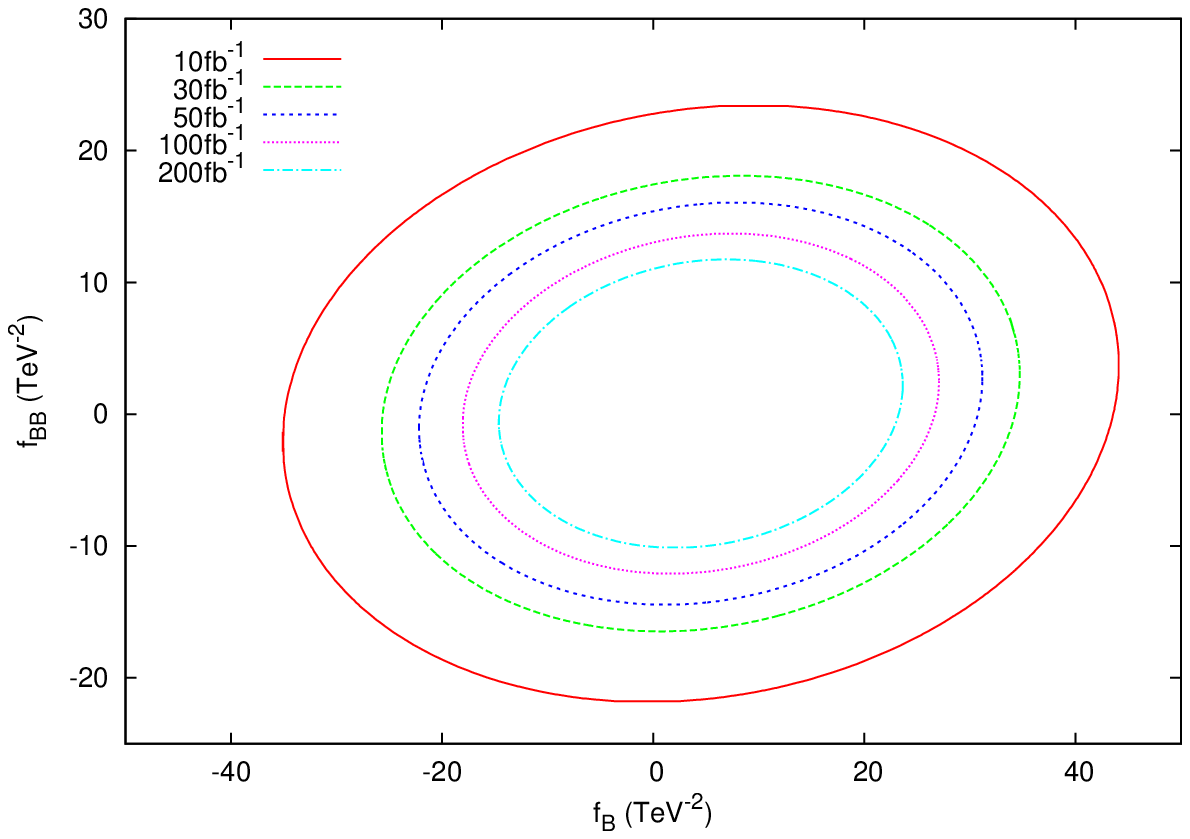}
\caption{The areas restricted by the lines show 95\% C.L.
sensitivity bounds on the parameter space $f_B-f_{BB}$ for various
integrated LHC luminosities stated on the figure. The scenario V is
taken into consideration. The center of mass energy of the
proton-proton system is taken to be $\sqrt s=14$TeV.\label{fig7}}
\end{figure}

\begin{figure}
\includegraphics[scale=1]{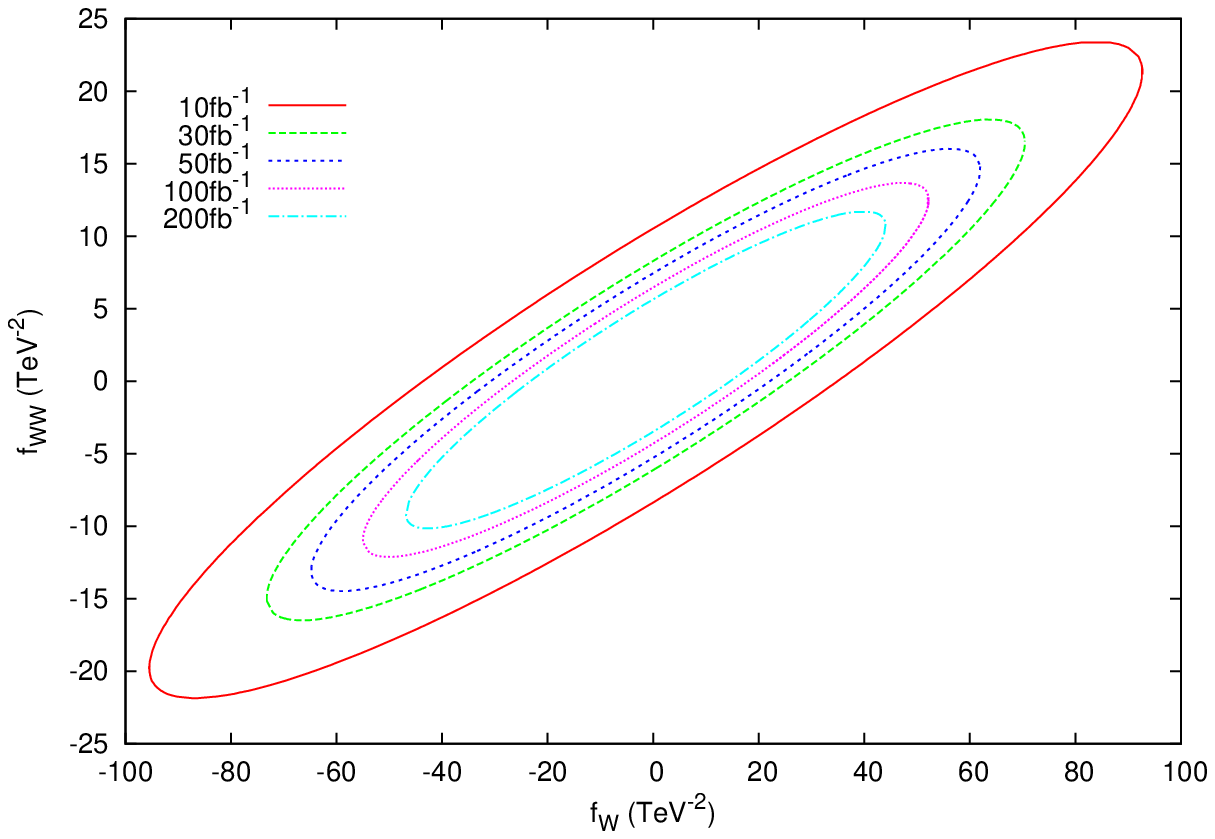}
\caption{The areas restricted by the lines show 95\% C.L.
sensitivity bounds on the parameter space $f_W-f_{WW}$ for various
integrated LHC luminosities stated on the figure. The scenario VI is
taken into consideration. The center of mass energy of the
proton-proton system is taken to be $\sqrt s=14$TeV.\label{fig8}}
\end{figure}

In Table \ref{tab1} we present 95\% C.L. bounds on non-standard
$f_{ww}$,$f_{w}$ and $f_{bb}$ couplings for scenarios I-IV. The
bounds are obtained via one-parameter $\chi^2$ analysis and we
consider the integrated luminosity values of
$L_{int}=10,30,50,100,200\;fb^{-1}$. For scenarios V and VI we have
two free coupling parameters and therefore the bounds are obtained
using two-parameter $\chi^2$ analysis. In Fig.\ref{fig7} and
Fig.\ref{fig8}, we plot 95\% C.L. bounds on two dimensional
parameter spaces $f_B-f_{BB}$ and $f_W-f_{WW}$ for scenarios V and
VI respectively.

The CMS collaboration at the LHC has determined direct experimental
bounds on non-standard Higgs-gauge boson couplings by studying Higgs
boson decay to $ZZ$, $Z\gamma$, $\gamma\gamma$ and $WW$
\cite{Khachatryan:2014kca}. The following 95\% C.L. bounds have been
given on the ratio of $HZ\gamma$ and $H\gamma\gamma$ couplings to
$HZZ$: $-0.046<\frac{a^{Z\gamma}_2}{a_1}<0.044$ and
$-0.011<\frac{a^{\gamma\gamma}_2}{a_1}<0.054$
\cite{Khachatryan:2014kca}. Here, $a$ couplings are defined by
$a_1=2g_{HZZ}/m_Z^2$, $a^{Z\gamma}_2=g^{(2)}_{HZ\gamma}$ and
$a^{\gamma\gamma}_2=2g_{H\gamma\gamma}$, where $g_{H\gamma\gamma}$
and $g^{(2)}_{HZ\gamma}$ are the couplings in the effective
lagrangian in Eq.(\ref{Leff2}) and $g_{HZZ}$ is the coupling of the
Higgs to two Z boson, i.e., $g_{HZZ}HZ^\mu Z_\mu$. If we assume that
$g_{HZZ}$ coupling is equal to its SM value ($g_{HZZ}=m_Z^2/\nu$;
$\nu=246$ GeV) then we can extract the experimental bounds on the
couplings $g^{(2)}_{HZ\gamma}$ and $g_{H\gamma\gamma}$. The scenario
III and scenario IV isolate the couplings $g^{(2)}_{HZ\gamma}$ and
$g_{H\gamma\gamma}$ respectively. Therefore, these scenarios give us
the opportunity to compare our bounds with the experimental bounds
of Ref.\cite{Khachatryan:2014kca}.  In scenario III, the
experimental bound on $\frac{a^{Z\gamma}_2}{a_1}$ can be converted
to the bounds on $f$ couplings as $-13<f_{BB}<12.6$ and
$-12.6<f_{WW}<13$. Similarly, in scenario IV the experimental bound
on $\frac{a^{\gamma\gamma}_2}{a_1}$ can be converted as
$-5.67<f_{BB}<27.83$ and $-1.7<f_{WW}<8.35$. When we compare these
bounds with the corresponding bounds given in Table \ref{tab1}, we
see that our bounds for the integrated luminosity of $200\; fb^{-1}$
are approximately a factor of 3 better than the experimental bounds
in the case of scenario III and approximately a factor of 2.5 better
in the case of scenario IV.


\section{Single Higgs production through photon-electron collision at the future linear collider}

\begin{figure}
\includegraphics[scale=1]{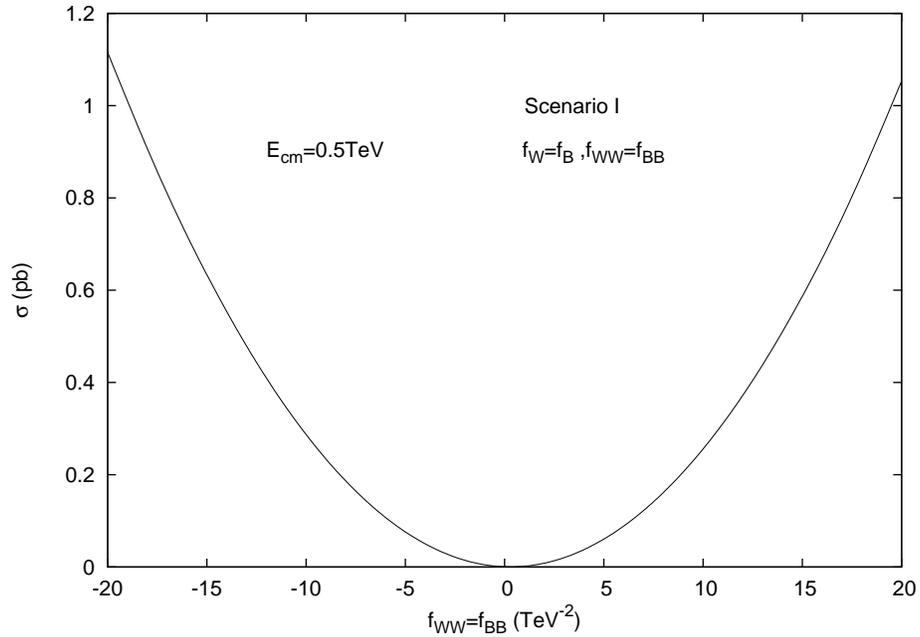}
\caption{The total cross section observed in the $e^-e^+$ collision
as a function of non-standard Higgs coupling for scenario I. The
main $e^-e^+$ collider energy is taken to be $\sqrt
s=0.5$TeV.\label{fig9}}
\end{figure}

\begin{figure}
\includegraphics[scale=1]{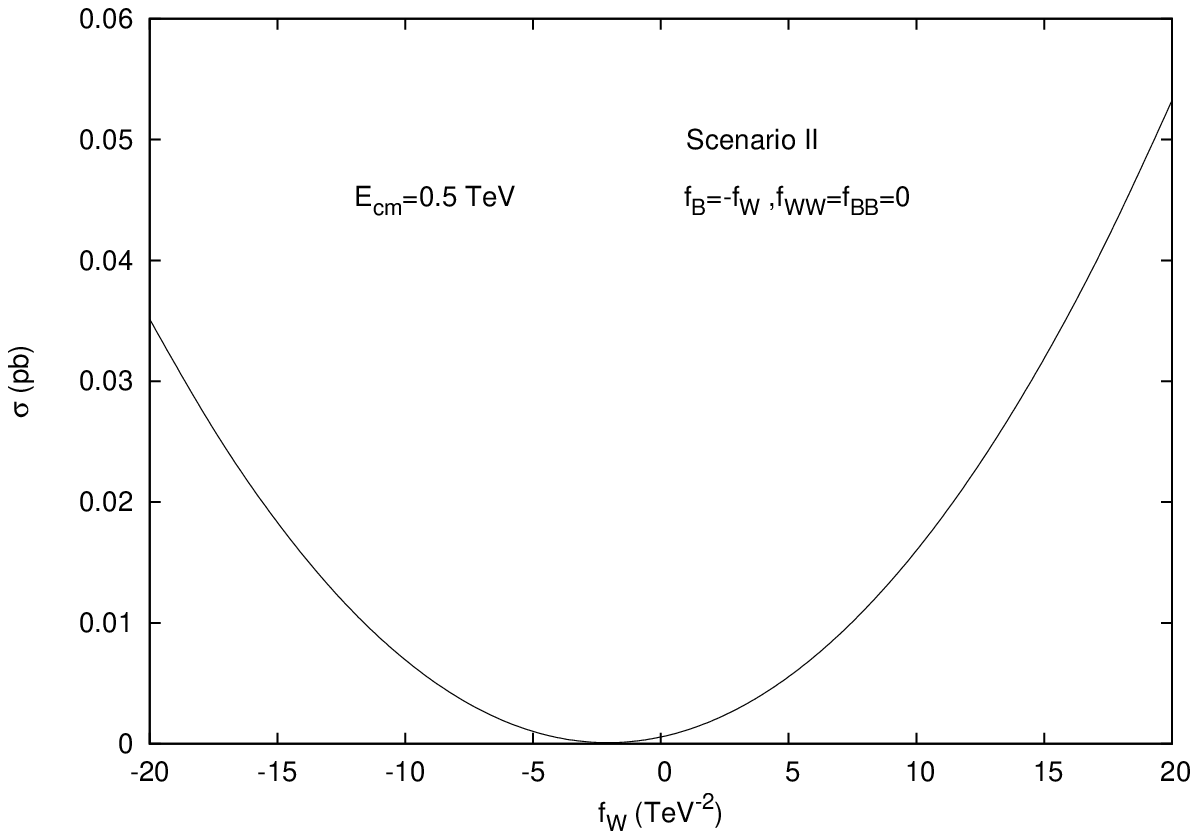}
\caption{The same as Fig.\ref{fig9} but for scenario
II.\label{fig10}}
\end{figure}

\begin{figure}
\includegraphics[scale=1]{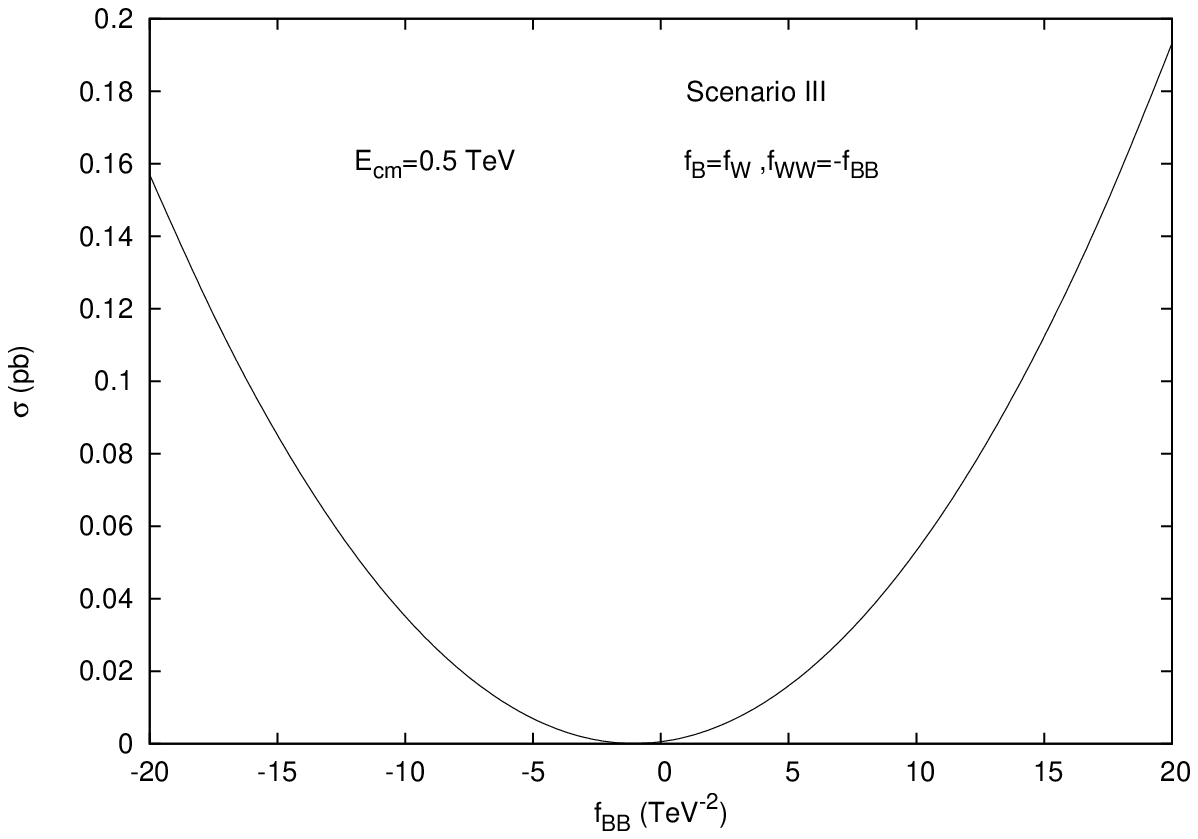}
\caption{The same as Fig.\ref{fig9} but for scenario
III.\label{fig11}}
\end{figure}

\begin{figure}
\includegraphics[scale=1]{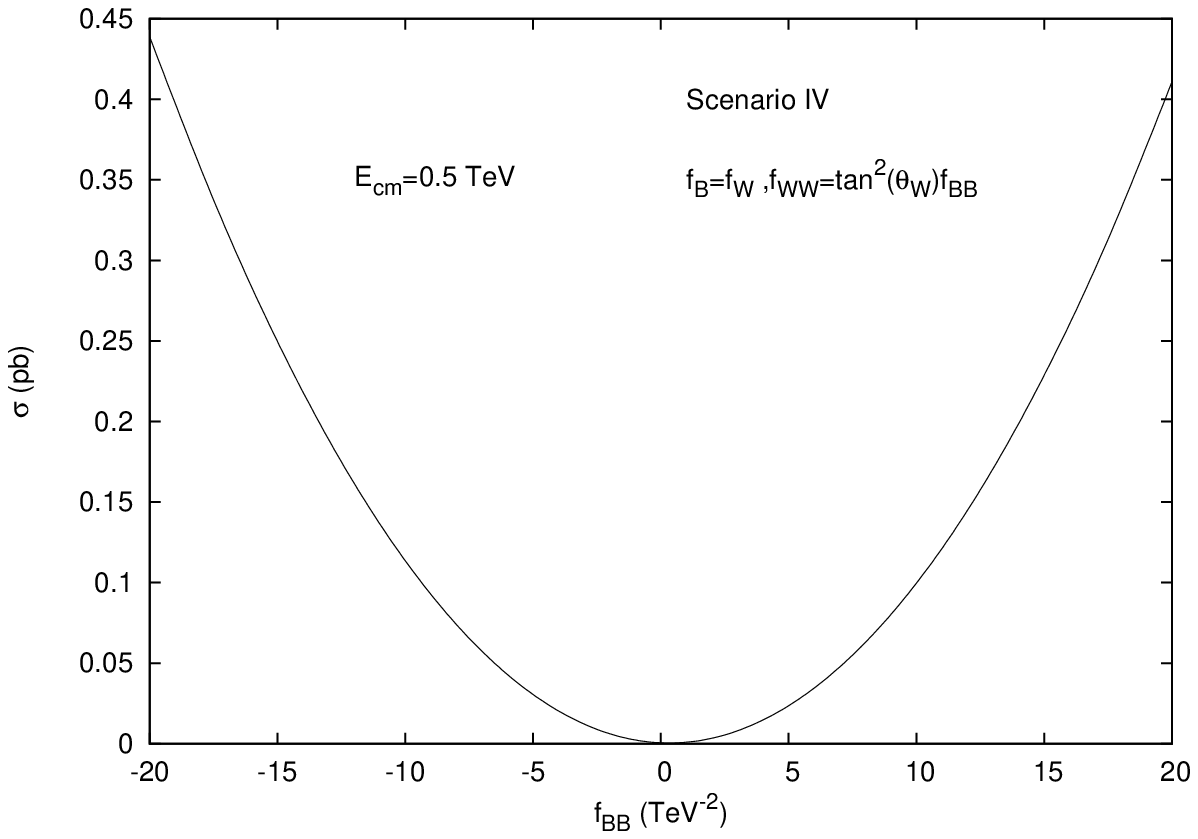}
\caption{The same as Fig.\ref{fig9} but for scenario
IV.\label{fig12}}
\end{figure}

The non-standard $H\gamma\gamma$ and $HZ\gamma$ couplings can be
investigated with a high precision at future linear $e^-e^+$
collider and its $e\gamma$ and $\gamma\gamma$ modes. We consider the
single Higgs production in the $e\gamma$ collision via the
subprocess $\gamma e\to He$. The tree-level Feynman diagrams for
$\gamma e\to He$ is very similar to that of Fig.\ref{fig1}, but we
should replace quarks with electrons (or positrons), $q\to e$. The
initial photon beam can be obtained through equivalent photon
emission from incoming electron or positron beam, similar to
equivalent photon emission from protons at the LHC. However in the
case of future linear collider, we have a more appealing option. A
real photon beam can be obtained through Compton backscattering of
laser light off the linear electron beam. Contrary to EPA, Compton
backscattering provides an increasing photon spectrum as a function
of the energy fraction $y=E_\gamma/E_e$, where $E_\gamma$ and $E_e$
represent the energy of the backscattered photon and initial
electron beam, respectively
\cite{ComptonBackScattering1,ComptonBackScattering2}. The
backscattered photon spectrum is given by
\cite{ComptonBackScattering1,ComptonBackScattering2}
\begin{eqnarray}
f_{\gamma/e}(y)={{1}\over{g(\zeta)}}[1-y+{{1}\over{1-y}}
-{{4y}\over{\zeta(1-y)}}+{{4y^{2}}\over {\zeta^{2}(1-y)^{2}}}]
\end{eqnarray}
where,
\begin{eqnarray}
g(\zeta)=&&(1-{{4}\over{\zeta}}
-{{8}\over{\zeta^{2}}})\ln{(\zeta+1)}
+{{1}\over{2}}+{{8}\over{\zeta}}-{{1}\over{2(\zeta+1)^{2}}} \;.
\end{eqnarray}
Here, $\zeta=4E_{e}E_{0}/M_{e}^{2}$ and $E_{0}$ is the energy of
initial laser photon before Compton backscattering. The $\zeta$
parameter can be taken to be $\zeta=4.8$ in which case the
backscattered photon energy is maximized without spoiling the
luminosity. Then, the upper limit of the energy fraction becomes
$y_{max}=0.83$. The process $\gamma e\to He$ takes part as a
subprocess in the main $e^-e^+$ collision. Therefore, the total
cross section observed in the $e^-e^+$ collision can be obtained by
integrating the cross section for $\gamma e\to He$ over the
backscattered photon spectrum:
\begin{eqnarray}
\label{integratedcrosssection} \sigma_{e^-e^+}=\int_{y_{min}}^{0.83}
f_{\gamma/e}(y)\; \sigma_{\gamma e\to He}\; dy
\end{eqnarray}
where, $y_{min}=\frac{m_{H}^{2}}{s}$ and $s$ is the Mandelstam
parameter of the $e^-e^+$ collision. The behavior of the total cross
section as a function of non-standard Higgs couplings is shown in
Figs.\ref{fig9}-\ref{fig12} for scenarios I-IV. In these figures,
the center of mass energy of the main $e^-e^+$ collider is taken to
be $\sqrt s=0.5$ TeV. Since the mass of the electron is very tiny
the tree-level SM contribution to the process $\gamma e\to He$ can
be safely neglected. Therefore the SM contributions to $\gamma e\to
He$ are coming from the loop-level. We consider SM one-loop
contributions described by the effective lagrangian (\ref{loop}).

\begin{figure}
\includegraphics[scale=1]{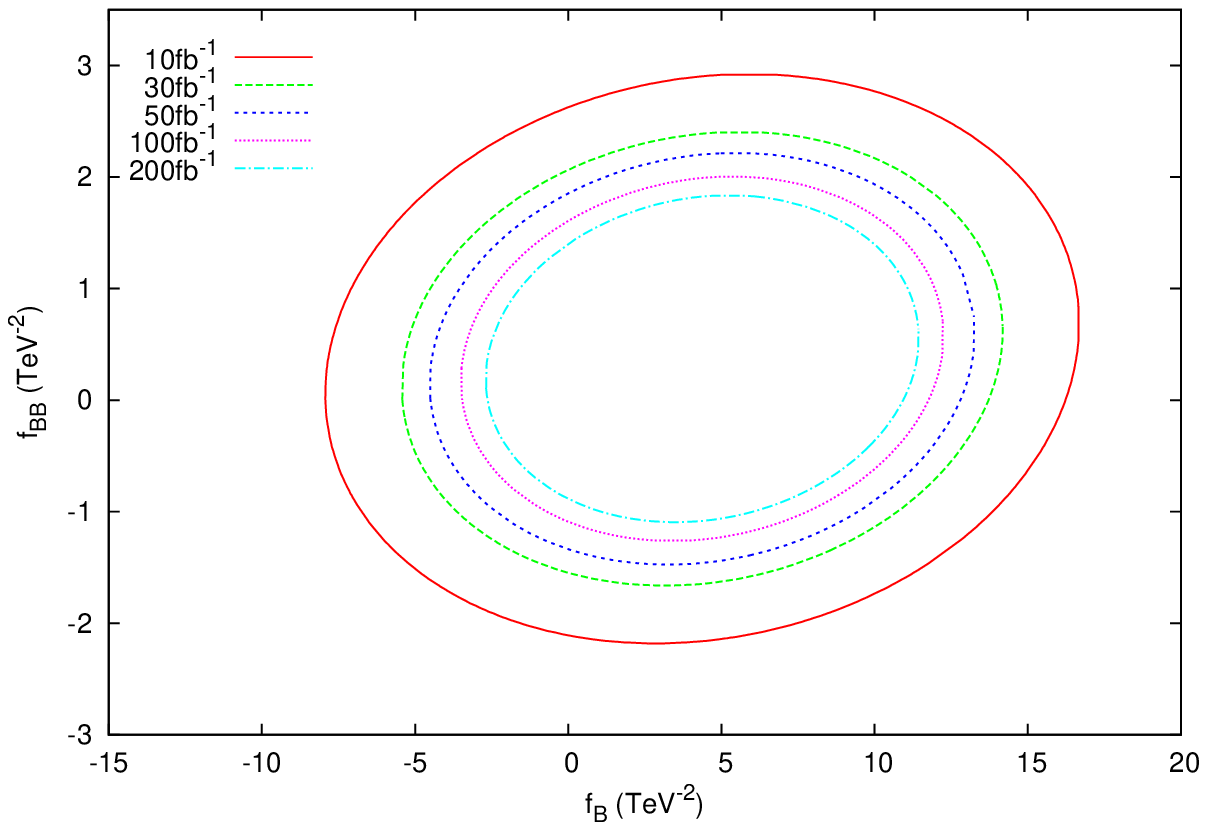}
\caption{The areas restricted by the lines show 95\% C.L.
sensitivity bounds on the parameter space $f_B-f_{BB}$ for various
integrated linear collider luminosities stated on the figure. The
scenario V is taken into consideration. The main $e^-e^+$ collider
energy is taken to be $\sqrt s=0.5$TeV.\label{fig13}}
\end{figure}

\begin{figure}
\includegraphics[scale=1]{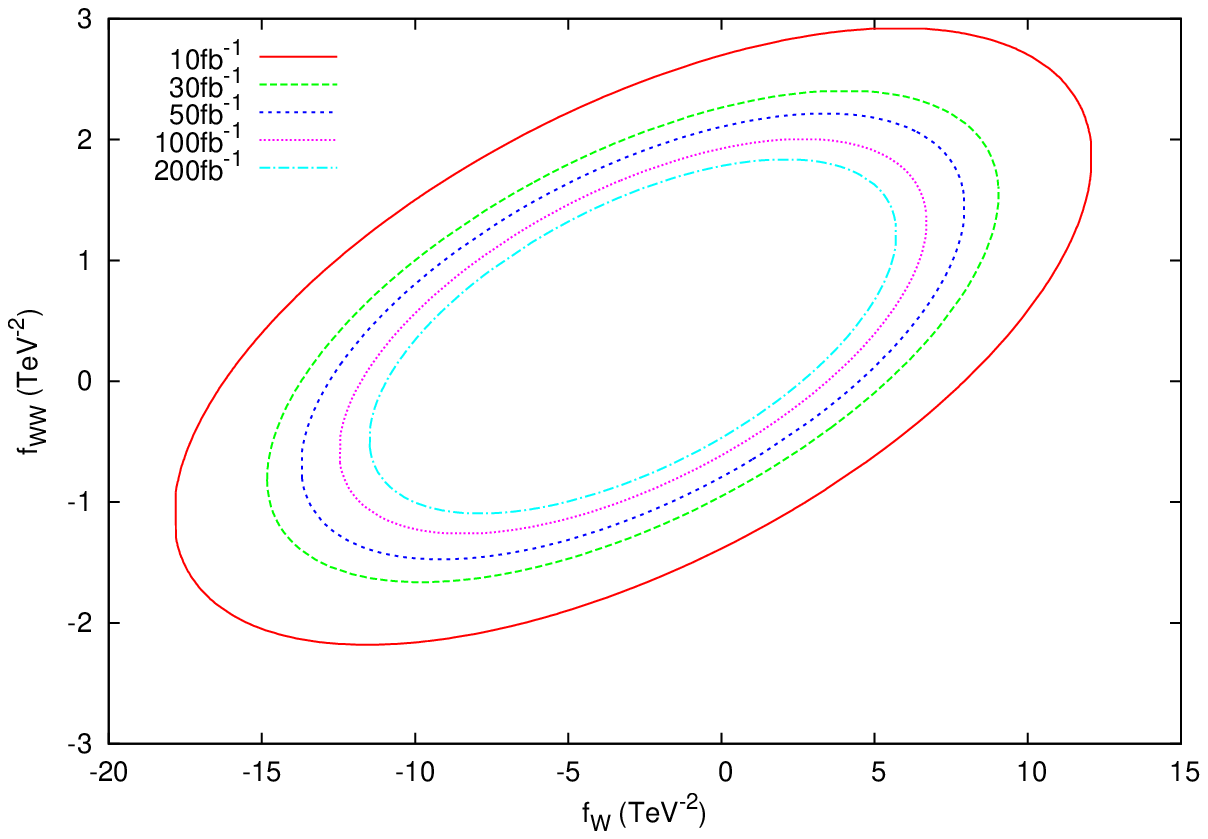}
\caption{The areas restricted by the lines show 95\% C.L.
sensitivity bounds on the parameter space $f_W-f_{WW}$ for various
integrated linear collider luminosities stated on the figure. The
scenario VI is taken into consideration. The main $e^-e^+$ collider
energy is taken to be $\sqrt s=0.5$TeV.\label{fig14}}
\end{figure}

Using the simple $\chi^2$ criterion we estimate sensitivity of the
linear collider-based $e\gamma$ collider to non-standard Higgs
couplings for the integrated luminosity values of
$L_{int}=10,30,50,100,200\;fb^{-1}$ and $\sqrt s=0.5$ TeV. We
consider $H\to b \bar b$ decay channel of the Higgs boson and assume
that $b \bar b$ final state with invariant mass in the interval
$120\;\text{GeV}<M_{b\bar b}<130\;\text{GeV}$ is identified as the
signal. In the $\chi^2$ function the number of events is given by
$N_{NS(SM)}=E\times L_{int} \times \sigma_{NS(SM)}\times BR$. We
take into account a b-tagging efficiency of $E=0.6$ and branching
ratio of $BR=0.6$. We assume that the central detectors have a
pseudorapidity coverage of $|\eta|<2.5$. Therefore, we place a cut
of $|\eta|<2.5$ for all final state particles. The potential
background process is $\gamma e\to b\bar b e$. It is described by 8
tree-level Feynman diagrams and gives a total cross section of
$\sigma_{\text{background}}=4.1\times10^{-3}\;\text{pb}$ after
imposing the cuts $120\;\text{GeV}<M_{b\bar b}<130\;\text{GeV}$ and
$|\eta|<2.5$. Similar to the statistical analysis performed in the
previous section, we assume that the background contribution cannot
be discerned from Higgs production. Therefore, during statistical
analysis we add the background contribution to the SM cross section
and assume that
$\sigma_{NS(SM)}={(\sigma_{e^-e^+}})_{NS(SM)}+\frac{1}{BR}\times\sigma_{\text{background}}$
where ${(\sigma_{e^-e^+}})_{NS(SM)}$ is the integrated cross section
defined in (\ref{integratedcrosssection}). The subscript $NS$
represents the cross section containing both new physics and SM
contributions and subscript $SM$ represents the SM cross section
alone. The 95\% C.L. bounds on non-standard $f_{ww}$,$f_{w}$ and
$f_{bb}$ couplings are given in Table \ref{tab2} for scenarios I-IV.
We observe from Tables \ref{tab1} and \ref{tab2} that the bounds of
Table \ref{tab2} are more restrictive with respect to the
corresponding bounds of Table \ref{tab1}. The average improvement
factors are approximately $6$ for scenario I, $3$ for scenarios II
and III and $8.5$ for scenario IV. For scenarios V and VI the bounds
are obtained in the two-dimensional parameter spaces $f_{B}-f_{BB}$
and $f_{W}-f_{WW}$. The 95\% C.L. restricted regions in these
parameter spaces are given in Fig.\ref{fig13} and Fig.\ref{fig14}.
When we compare the bounds of Figs.\ref{fig13} and \ref{fig14} with
the similar LHC bounds given in Figs.\ref{fig7} and \ref{fig8}, we
see that the bounds of the linear collider are approximately a
factor of $5$ better than the corresponding bounds of the LHC.

We can also compare the bounds of future linear collider with the
current experimental bounds. The CMS bounds on $f_{BB}$ and $f_{WW}$
couplings have been given in the last paragraph of the previous
section. When we compare these experimental bounds with the
corresponding bounds given in Table \ref{tab2}, we see that our
bounds for the integrated luminosity of $200\; fb^{-1}$ are
approximately a factor of 8 better than the experimental bounds in
the case of scenario III and approximately a factor of 20 better in
the case of scenario IV.

\section{Conclusions}

One of the prominent motivations of the future $e^-e^+$ collider is
that it provides clean experimental environment which allows to make
high precision measurements \cite{Behnke:2013xla,Baer:2013cma}. In
deep inelastic hadron-hadron collisions, initial hadron beams
dissociate into partons and create myriad of jets which cause
uncertainties and make it difficult to discern the signals that we
want to observe. Moreover, in hadron colliders there are systematic
uncertainties arising from the proton structure functions, from
unknown higherorder perturbative QCD corrections, and from
nonperturbative QCD effects \cite{Baer:2013cma}. Lepton colliders do
not suffer from these kind of uncertainties, and the level of
precision is expected to be enhanced considerably compared to hadron
colliders. On the other hand, ultraperipheral collisions in a hadron
collider provides a unique opportunity to search for the physics
beyond the SM in a rather clean environment with respect to deep
inelastic hadron-hadron collisions. Exclusive and semielastic
processes are examples of the reactions in an ultraperipheral
collision. In semi-elastic Higgs production $pp\to p\gamma p\to
pHqX$ only one of the incoming proton dissociates into partons but
the other proton remains intact. The absence of the remnants of one
of the proton beam, allows to discern the signal more easily.
Furthermore, tagging the intact scattered protons in the forward
detectors allows us to reconstruct quasireal photons' momenta. The
knowledge obtained in this way is very useful in reconstructing the
kinematics of the reaction. The semi-elastic Higgs production is
electroweak in nature and free from backgrounds containing strong
interaction. Due to above reasons, the uncertainties associated with
the Higgs detection for $pp\to p\gamma p\to pHqX$ are expected to be
reduced considerably compared to deep inelastic processes at the
LHC. Therefore, the comparison of the results obtained in
semi-elastic production at the LHC and future $e^-e^+$ collider is
important and contributes to the physics program of the future
$e^-e^+$ collider.

In the paper, we consider similar subprocesses $\gamma q\to H q$ and
$\gamma e\to H e$ at the LHC and at future $e^-e^+$ collider. We
investigate the potential of these two colliders to probe
non-standard Higgs couplings. We show that $e\gamma$ mode of the
linear collider with center of mass energy of $\sqrt s=0.5$ TeV
probes the non-standard $HZ\gamma$ and $H\gamma\gamma$ couplings
with better sensitivity than the $\gamma$-proton collision at the
LHC. The improvement factor depends on the coupling and the
luminosity, but roughly the bounds are improved by a factor of 5.

\newpage

\begin{table}
\caption{95\% C.L. bounds on $f_{ww}$,$f_{w}$ and $f_{bb}$  for
various integrated LHC luminosities and scenarios. Bounds are given
in units of $\text{TeV}^{-2}$. The center of mass energy of the
proton-proton system is taken to be $\sqrt s=14$TeV.\label{tab1}}
\begin{ruledtabular}
\begin{tabular}{ccccc}

Luminosity&($Scenario$-I)%
$f_{ww}$&($Scenario$-II)$f_{w}$&($Scenario$-III)$f_{bb}$&($Scenario$-IV)$f_{bb}$
\\
\hline
10$fb^{-1}$ &(-6.3,7.9) &(-19.8,15.4) &(-9.9,7.7) &(-13.2,15.6) \\
30$fb^{-1}$ &(-4.6,6.2) &(-15.6,11.3) &(-7.8,5.6) &(-9.8,12.2) \\
50$fb^{-1}$ &(-3.9,5.6) &(-14.1,9.7) &(-7.0,4.9) &(-8.5,10.8) \\
100$fb^{-1}$ &(-3.2,4.8) &(-12.2,7.9) &(-6.1,3.9) &(-7.0,9.3) \\
200$fb^{-1}$ &(-2.6,4.2) &(-10.7,6.4) &(-5.3,3.2) &(-5.7,8.1) \\
\end{tabular}
\end{ruledtabular}
\end{table}

\begin{table}
\caption{95\% C.L. bounds on $f_{ww}$,$f_{w}$ and $f_{bb}$  for
various integrated linear collider luminosities and scenarios.
Bounds are given in units of $TeV^{-2}$. The main $e^-e^+$ collider
energy is taken to be $\sqrt s=0.5$TeV. \label{tab2}}
\begin{ruledtabular}
\begin{tabular}{ccccc}

Luminosity&($Scenario$-I)%
$f_{ww}$&($Scenario$-II)$f_{w}$&($Scenario$-III)$f_{bb}$&($Scenario$-IV)$f_{bb}$
\\
\hline
10$fb^{-1}$ &(-0.8,1.3) &(-7.6,3.4) &(-3.8,1.7) &(-1.3,2.0) \\
30$fb^{-1}$ &(-0.5,1.1) &(-6.5,2.3) &(-3.2,1.2) &(-1.0,1.6) \\
50$fb^{-1}$ &(-0.5,1.0) &(-6.1,1.9) &(-3.0,1.0) &(-0.8,1.5) \\
100$fb^{-1}$ &(-0.4,0.9) &(-5.6,1.5) &(-2.8,0.7) &(-0.6,1.3) \\
200$fb^{-1}$ &(-0.3,0.9) &(-5.3,1.1) &(-2.6,0.6) &(-0.5,1.2) \\
\end{tabular}
\end{ruledtabular}
\end{table}

\end{document}